\RequirePackage[running]{lineno}

\documentclass[aps,prl,twocolumn,nofootinbib,superscriptaddress,letterpaper,amsmath,amssymb]{revtex4}

\usepackage{graphicx}  
\usepackage{ulem}
\usepackage{dcolumn}   
\usepackage{bbm}        
\usepackage{amsmath}

\usepackage{amsfonts}
\usepackage{amssymb}   
\usepackage{epstopdf}
\usepackage{slashed}
\usepackage{hyperref}
\usepackage{multirow}
\usepackage{color}
\usepackage{float}
\usepackage{subcaption}
\usepackage{cleveref}

\usepackage{verbatim}

\usepackage[compat=1.1.0]{tikz-feynman}

\def\gsim{\lower0.5ex\hbox{$\:\buildrel >\over\sim\:$}}
\def\lsim{\lower0.5ex\hbox{$\:\buildrel <\over\sim\:$}}

\newcommand{\be}{\begin{equation}}
\newcommand{\ee}{\end{equation}}
\newcommand{\bea}{\begin{eqnarray}}
\newcommand{\eea}{\end{eqnarray}}

\newcommand{\del}{\partial}

\newcommand{\nbox}{{\,\lower0.9pt\vbox{\hrule \hbox{\vrule height 0.2 cm
\hskip 0.2 cm \vrule height 0.2 cm}\hrule}\,}}

\def\sub#1{_{\lower.25ex\hbox{$\scriptstyle#1$}}}

\newskip\zatskip \zatskip=0pt plus0pt minus0pt
\def\matth{\mathsurround=0pt}
\def\lsim{\mathrel{\mathpalette\atversim<}}
\def\gsim{\mathrel{\mathpalette\atversim>}}
\def\sigv{\ifmmode \langle\sigma v\rangle\else $\langle\sigma v\rangle$\fi}
\newskip\zatskip \zatskip=0pt plus0pt minus0pt
\def\matth{\mathsurround=0pt}
\def\lsim{\mathrel{\mathpalette\atversim<}}
\def\gsim{\mathrel{\mathpalette\atversim>}}
\def\atversim#1#2{\lower0.7ex\vbox{\baselineskip\zatskip\lineskip\zatskip
  \lineskiplimit
  0pt\ialign{$\matth#1\hfil##\hfil$\crcr#2\crcr\sim\crcr}}}

\begin{document}

\thispagestyle{empty}

\vspace{0.5in}

\title{Untangling New Physics in Single Resonant Top Quarks}
\author{Krish Wu}
\affiliation{The four first authors contributed equally}
\affiliation{University High School, Irvine, CA}
\author{Brandon Sun}
\affiliation{The four first authors contributed equally}
\affiliation{Department of Information \& Computer Sciences, University of California, Irvine}
\author{Nitish Polishetty}
\affiliation{The four first authors contributed equally}
\affiliation{Vista Ridge High School, Cedar Park, Texas}
\author{Justin Kline}
\affiliation{The four first authors contributed equally}
\affiliation{Sage Hill School, Newport Beach, CA}
\author{Max Fieg}
\author{Daniel Whiteson}
\affiliation{Department of Physics \& Astronomy, University of California, Irvine}
\begin{abstract}
Collisions of particles at the energy frontier can reveal new particles and forces via localized excesses. However, the initial observation may be consistent with a large variety of theoretical models, especially in sectors with new top quark partners, which feature a rich set of possible underlying interactions.  We explore the power of the LHC dataset to distinguish between models of the singly produced heavy top-like quark which interacts with the Standard Model through an electromagnetic form factor. We study the heavy top decay to a top quark and a virtual photon which produces a pair of fermions, propose a technique to disentangle the models, and calculate the expected statistical significance to distinguish between various hypotheses.
\end{abstract}
\maketitle

\section{Introduction}

Many theories of new physics predict striking phenomena observable in high energy collisions.  In some cases, the potential experimental signatures are consistent with several distinct models of new particles or interactions, and untangling them can be quite difficult~\cite{Arkani-Hamed:2005qjb,Alwall:2008ag}. The  discovery of the Higgs boson~\cite{ATLAS:2012yve,CMS:2012qbp}, for example, has been followed by precision measurements of its properties~\cite{ATLAS:2022tnm,CMS:2021kom} to determine whether it is the garden variety boson of the Standard Model, or something more exotic, a question whose answer is still being resolved.  In the event that a discovery of a new particle is made, whether anticipated or not~\cite{Craig:2016rqv,Kim:2019rhy}, it can be expected to spark a similar effort to distinguish potential theoretical explanations. The top quark sector has a rich set of models predicting new heavy partners, with multiple experimental signatures. In some cases, as in the scenario studied here, the nature of the production and decay can be quite similar across theoretical models, even if the form of the interactions is quite different.  

Recently, we proposed~\cite{Tong:2023lms} the study of single production of a heavy vector-like quark $T$, which for high $T$ masses features higher cross sections than pair production, and can include decays such as $T\rightarrow t\gamma$~\cite{Kim:2018mks}, which can dominate in some scenarios~\cite{Alhazmi:2018whk} over the more commonly studied $T\rightarrow Wb,Zt, Ht$. Searches by the LHC collaboration have primarily studied $T$ decays featuring heavy  bosons, $T\rightarrow Wb,Zt, Ht$~\cite{ATLAS:2018ziw,ATLAS:2017nap,CMS:2023agg,CMS:2022yxp,CMS:2020ttz,Roy:2020fqf,ATLAS:2022ozf,ATLAS:2023pja,ATLAS:2022tla}. Studies of $T\rightarrow \gamma t$ have received less attention, and  primarily focused on pair production of the new heavy fermion~\cite{Alhazmi:2018whk}.  ATLAS explored the $t\gamma$ final state~\cite{ATLAS:2023qdu}, but did not search for a $t\gamma$ resonance.

In this paper, we revisit single production of heavy vector-like quarks that interact through the photon via electromagnetic (EM) form factor operators. We study again the single production mode and focus on the decay $T\rightarrow t\gamma\rightarrow t f^+ f^-$, where the photon is off-shell and converts to a pair of fermions ($f^+ f^-$). This mode is shared by the dimension-5 and dimension-6 EM form factor operators, which allows us to explore several distinct theoretical models, predict $T$ production at the LHC, and examine the power of the LHC dataset to distinguish between these models.
    
The paper is organized as follows. The first section describes the theoretical models of the heavy top partner. The next section details our study of the experimental sensitivity to the various models. The following two sections discuss the results and present the conclusions, respectively.

\section{Models}\label{sec:model}

Vector-like top partners~\cite{Alves:2023ufm} are typically found in models that alleviate the hierarchy problem and are less constrained than a 4th generation of quarks with chiral couplings \cite{Aguilar_Saavedra_2013,Dobrescu:1997nm,duPlessis:2021xuc}. These particles can be found in more complete models such as Little Higgs models \cite{Schmaltz:2005ky}, supersymmetric theories, composite Higgs models \cite{Witzel:2019jbe} or in scenarios where the right-handed top quark is composite and $T$ is an excited state\cite{Lillie:2007hd,Delgado:2005fq}. However, the compatibility of these motivations and the model we study are dependent on the details of the underlying theory.

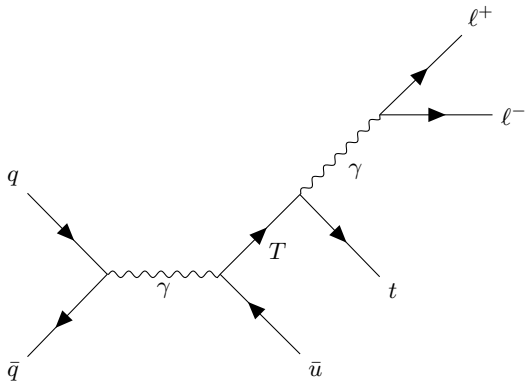
\begin{figure}[h!]
    \centering
      \begin{tikzpicture}
    \begin{feynman}
        \vertex (a) ;
        \vertex [above left=of a](i1){\(q\)};
        \vertex [below left=of a](i2){\(\bar{q}\)};
        \vertex [right=of a] (b);
        \vertex [above right=of b] (f1); 
        \vertex [above right=of f1] (fa);
        \vertex [below right=of f1] (fb) {\(t\)};
        \vertex [above right=of fa] (ffa) {\(\ell^+\)};
        \vertex [right=of fa] (ffb) {\(\ell^-\)};
        \vertex [below right=of b] (c) {\(\bar{u}\)};
        \diagram* {
         (i1) -- [fermion] (a),
         (i2) -- [anti fermion] (a),
        (a) -- [photon, edge label'=\(\gamma\)] (b) -- [fermion, edge label'=\(T\)] (f1),
         (f1) -- [photon, edge label'=\(\gamma\)] (fa),
         (f1) -- [fermion] (fb),
         (fa) -- [fermion] (ffb),
         (fa) -- [fermion] (ffa),
        (b) -- [anti fermion] (c),
    };
    \end{feynman}
    \end{tikzpicture}
    \caption{Feynman diagram describing the single production of a heavy vector-like fermionic top partner $T$ which decays to a top quark ($t$) and a photon ($\gamma$), which then decays to a pair of charged leptons ($\ell^+\ell^-$). 
    }
    \label{fig:decay-diagram}
\end{figure}

In this work, we consider a class of models where a vector-like quark $T$, can interact with the SM up-type quarks through electromagnetic form factor operators up to dimension-6. In particular, we consider an interaction taking the form of a dimension-5 magnetic and electric dipole moment (MDM, EDM), and the dimension-6 anapole and charge radius (CR) operators. The Lagrangians that govern these interactions can be written as the following:

\begin{align}
{\cal L}_{\rm MDM} &= \mu_i ~\bar{T}\sigma^{\mu \nu}u_{R,i}F_{\mu\nu} , \label{eq:lagrangianMDM}\\
{\cal L}_{\rm EDM} &= d_i ~\bar{T}\sigma^{\mu \nu}\gamma^5 u_{R,i}F_{\mu\nu}, \label{eq:lagrangianEDM}\\
{\cal L}_{\rm Anapole} &= a_i ~\bar{T}\gamma^{\mu}\gamma^5 u_{R,i}\del^{\nu}F_{\mu\nu}, \label{eq:lagrangianAnapole}\\
{\cal L}_{\rm CR} &= b_i~ ~\bar{T}\gamma^{\mu} u_{R,i}\del^{\nu}F_{\mu\nu} ,\label{eq:lagrangianCR}
\end{align}

\noindent where $u_{R,i}$ is a right-handed up-type quark, $\mu_i,d_i,a_i,b_i$ are dimensionful couplings of $T$ to a quark with flavor $i$. 

The dimension-5 operators induce a $T\rightarrow t \gamma$ decay,
and while the partial decay width of this channel for the dimension-6 operators is zero for an on-shell photon, the offshell photon decaying to a pair of fermions has non-zero width. Thus, the four models produce $t f^+ f^-$, allowing a direct comparison of the kinematics in the same final states.

\section{Experimental Sensitivity}
\label{sec:exp}

The models described above include interactions which can generate a final state with a top quark and a pair of opposite-sign charged leptons; see Fig.~\ref{fig:decay-diagram}. We estimate the sensitivity of the LHC dataset to these hypothetical signals using samples of simulated $pp$ collisions at \mbox{$\sqrt{s}=13$ TeV} with an integrated luminosity of 300 fb$^{-1}$.

Samples of simulated signal and background events are used to model the reconstruction of the $T$ quark candidates, estimate selection efficiencies and expected signal and background yields. Collisions and decays are simulated with {\sc Madgraph5} v3.5.7 ~\cite{madgraph} with {\sc nnpdf} v2.3~\cite{NNPDF:2021uiq} and renormalization and factorization scales set to $m_Z$, and {\sc Pythia} v8.306~\cite{pythia} is used for fragmentation and hadronization. Radiation of additional gluons is modeled by {\sc Pythia}. The detector response is simulated with {\sc Delphes} v3.5.0~\cite{delphes} using the standard CMS card, extended to include an additional reconstruction of wide-cone jets, and {\sc root} version 5.34.25 \cite{ROOT}. 

Selected isolated  photons and leptons are required to have transverse momentum $p_\textrm{T}\geq10$ GeV and absolute pseudo-rapidity $0\leq|\eta|\leq2.5$. Isolation requires that less than 12\% (25\%) of the $p_\textrm{T}$ of the electron or photon (muon) be deposited in a cone with $\Delta R < 0.5$ centered on the particle. Selected narrow-cone (wide-cone) jets are clustered using the anti-$k_{\textrm{T}}$ algorithm~\cite{Cacciari:2008gp} with radius parameter $R = 0.4$ ($R=1.2$) using \textsc{FastJet 3.1.2}~\cite{Cacciari:2011ma} and are required to have $p_\textrm{T}\geq20$ GeV and $0\leq|\eta|\leq2.5$. Wide-cone jets with mass within $[50,110]$ ($[125,225]$) GeV are tagged as $W$-boson (top-quark) jets. Events are required to have exactly two opposite-sign electrons or muons, which are combined to reconstruct the photon candidate.

Reconstruction of the top quark can proceed in several ways due to the various decay modes of the $W$ boson.  The hadronic decay mode is typically the most powerful~\cite{CMS:2018rkg,ATLAS:2020lks} due to its high branching fraction and techniques to efficiently reconstruct its decay using large-radius jets and state-of-the-art tagging algorithms. The background from multijet events can be accurately modeled using data driven methods. Candidate $T$ quarks are reconstructed from the combination of the off-shell photon candidate (via the observed lepton pair) and the top quark candidate, which is reconstructed in one of three approaches: 

\begin{itemize}
    \item $t$: one top-tagged, $b$-tagged wide-cone jet 
    \item $W+b$: one $W$-tagged, un-$b$-tagged wide-cone jet and one $b$-tagged narrow-cone jet
    \item $jj+b$: two un-$b$-tagged narrow-cone jets and one $b$-tagged narrow-cone jet
\end{itemize}

\begin{figure}
    \centering
    \includegraphics[width=0.45\textwidth]{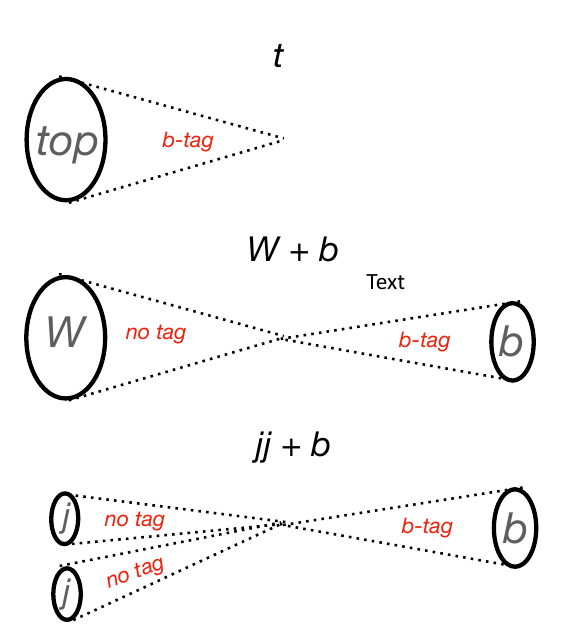}
    \caption{Three possible top quark  reconstruction strategies, using wide-cone and narrow-cone jets, which can be $b$-tagged, $W$-tagged or top-tagged. See text for details.}
    \label{fig:topo}
\end{figure}

\begin{figure}
    \centering
    \includegraphics[width=0.45\textwidth]{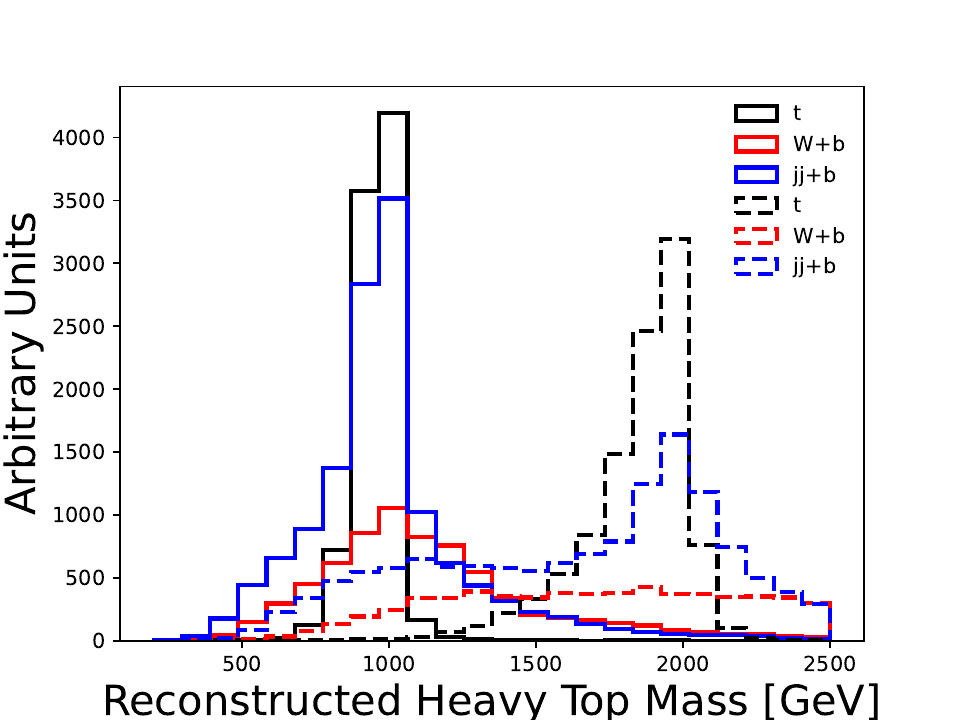}
    \includegraphics[width=0.45\textwidth]{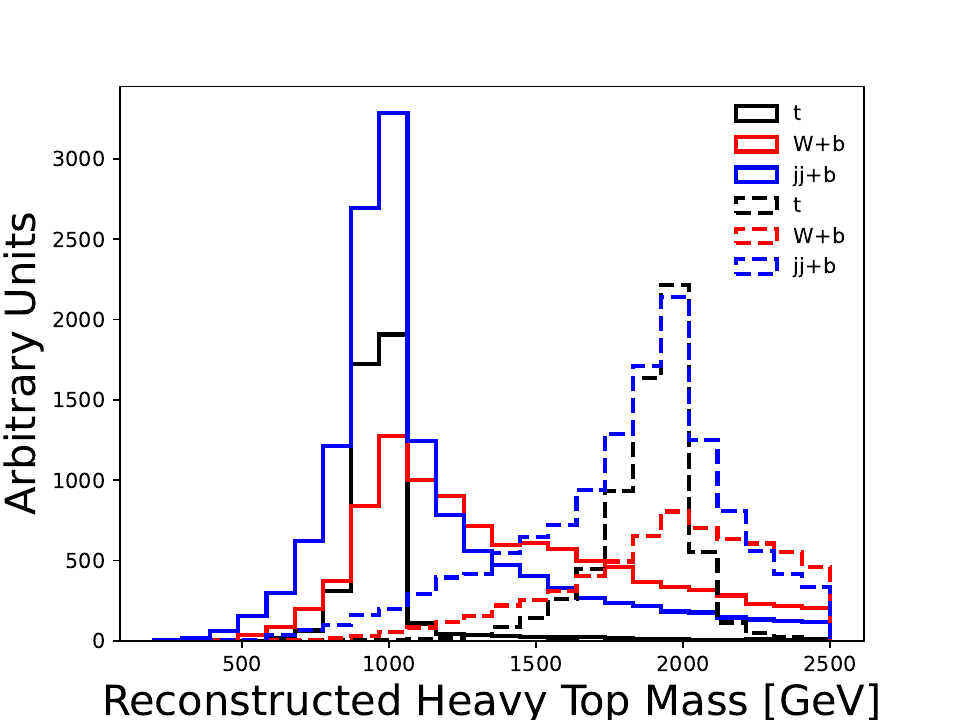}
    \caption{ Distribution of the reconstructed heav top ($T$) quark candidate mass in simulated events with $m_T=1000$ (solid) or 2000 (dashed) GeV, for MDM (top) and Anapole (bottom) models for each of the three reconstruction strategies (see Fig~\ref{fig:topo}). The overall normalization is arbitrary, but the relative density of reconstruction modes reflects the different top quark $p_\textrm{T}$ predicted by the two models in the $T\rightarrow t\gamma$ decay.} 
    \label{fig:reco}
\end{figure}

The three approaches are illustrated in Fig.~\ref{fig:topo}. If several reconstruction approaches are available for a single event, preference is given to $t$ and then $W+b$. If several jets are available within one approach, preference is given to the jets that minimize the difference between the reconstructed and known top-quark and $W$-boson masses.  Events with no $T$ quark candidate with mass greater than 200 GeV are rejected. Distributions of reconstructed $T$ quark candidate masses are shown in Fig.~\ref{fig:reco} for two representative models. 

\begin{figure}
    \centering
        \includegraphics[width=0.45\textwidth]{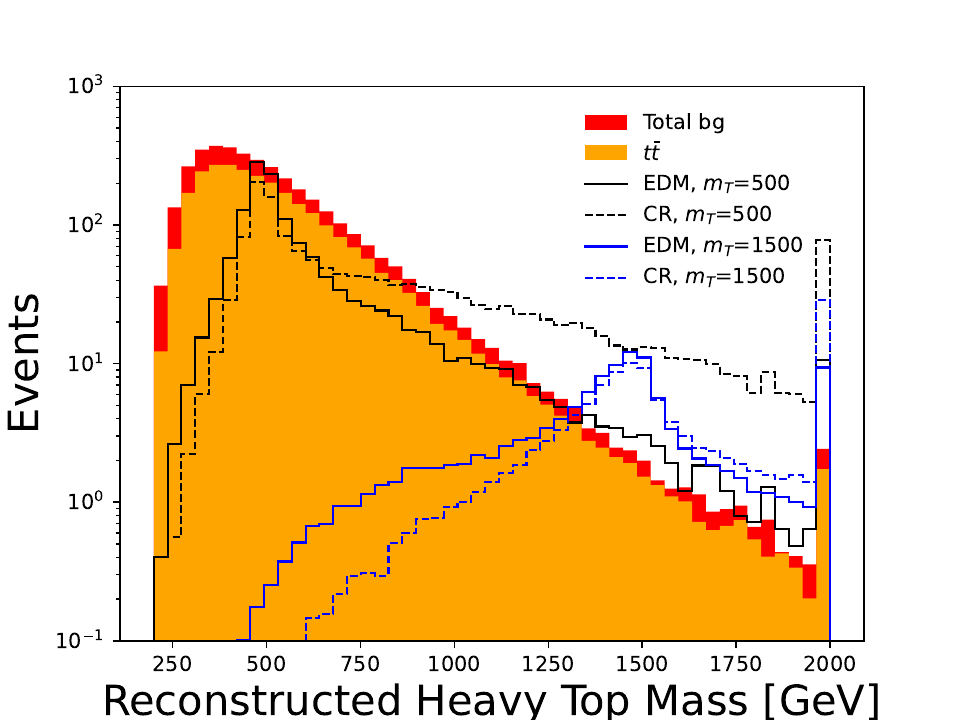}
                \includegraphics[width=0.45\textwidth]{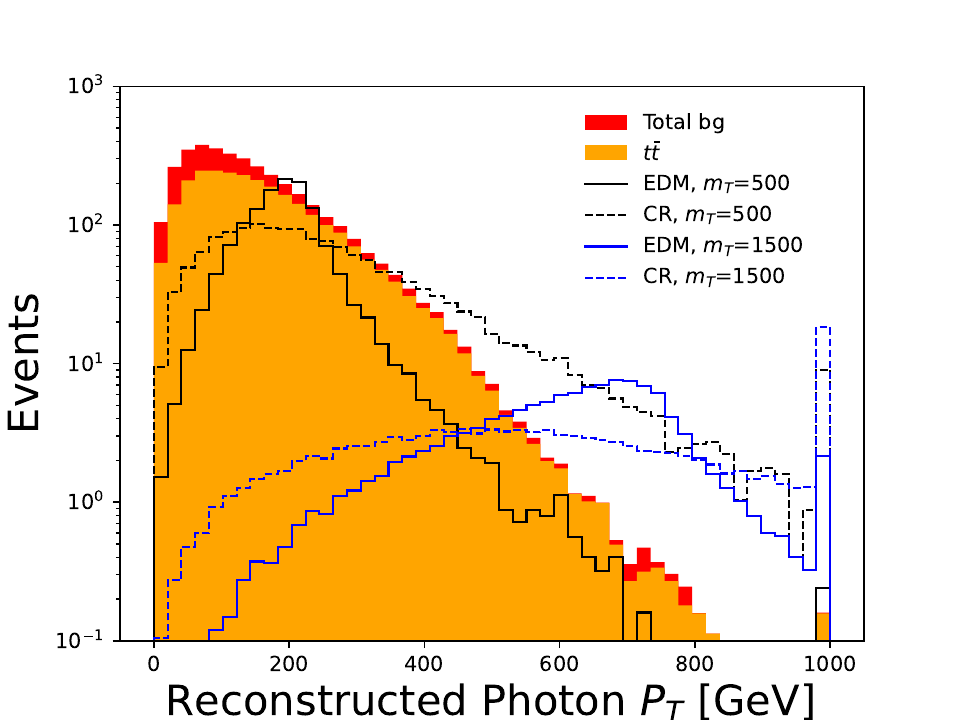}
                  \caption{                   Distribution of reconstructed heavy quark invariant mass (top) or photon $p_\textrm{T}$ in simulated signal and background samples,  normalized to an integrated luminosity of 300 fb$^{-1}$. The last bin shows overflow. The signal cross section is set to the expected limit at 95\% CL.} 
    \label{fig:sb}
\end{figure}

The dominant backgrounds are the production of top-quark pairs with an additional pair of leptons ($t\bar{t}\ell^+\ell^-$), or the production of a single top quark in association with a $b$ quark ($t\bar{b}\ell^+\ell^-$ or $\bar{t}b\ell^+\ell^-$) and a pair of leptons. Additional backgrounds are due to production of a heavy vector boson ($W$ or $Z$ bosons) in association with $b$ quarks and a pair of leptons, $W^\pm b\bar{b}\ell^+\ell^-$ or $Zb\bar{b}\ell^+\ell^-$.  Contribution from QCD multi-jet production is suppressed by the lepton requirement and the minimum $m_T$, and are  non-zero but sub-dominant and are neglected here. Distributions of the expected reconstructed $T$ quark masses and the photon transverse momentum for the background and signal processes are shown in Fig.~\ref{fig:sb} and the expected background yields in 300 fb$^{-1}$ are shown in Table~\ref{tab:yields}. 

\begin{table}[]
    \centering
     \caption{Expected yields ($N$) in 300 fb$^{-1}$ of LHC $pp$ collisions at $\sqrt{s}=13$ TeV for several background  processes. Cross sections for backgrounds are at NLO in QCD~\cite{Alwall:2014hca}.}
    \label{tab:yields}
    \begin{tabular}{l|ccc}
    \hline\hline
    Process &  $N$\\
    \hline
    $t\bar{t}\gamma$     & 2730 \\
    $Z+b\bar{b}+\gamma$	&681 \\
        $W+b\bar{b}+\gamma$&	256 \\
    $t\bar{b}\gamma+\bar{t}b\gamma$& 21 \\
         \hline\hline
    \end{tabular}
   
\end{table}

\subsection{Statistical Analysis}
\begin{figure}
     \centering
     \vspace{3cm}
     \scalebox{0.45}{\includegraphics{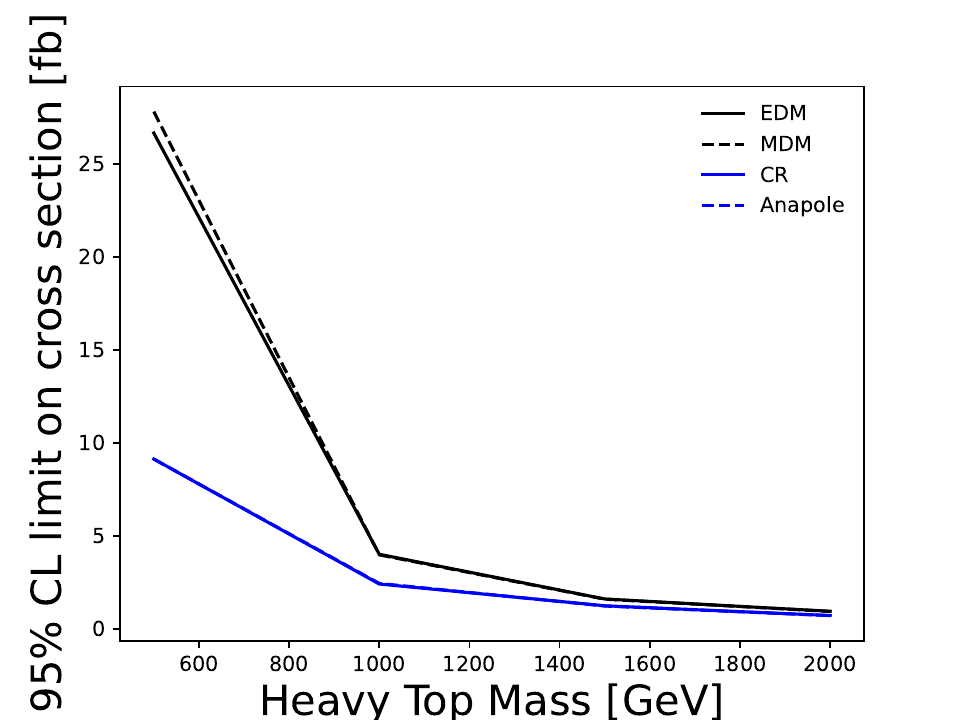}}
     \caption{ Expected 95\% CL upper limits on the production cross section for signal models, as a function of heavy top quark mass.}
     \label{fig:xs_limit_summary}
 \end{figure}

Expected limits on the single $T$ production cross section are calculated at 95\% confidence level (CL) using a profile likelihood ratio~\cite{Cowan:2010js}  with the CLs technique~\cite{Junk:1999kv,Read:2002hq}. We use the pyhf~\cite{pyhf, pyhf_joss} package with a binned distribution in heavy quark reconstructed mass, where bins without simulated background events have been merged into adjacent bins. The background is assumed to have a 50\% relative systematic uncertainty.   Limits on the signal cross section as a function of the $T$ quark mass are shown in Fig.~\ref{fig:xs_limit_summary}. 

Distinguishing between the various models using only the reconstructed heavy quark invariant mass would be challenging given the similarity in the mass distributions. But there is more information present in the events that can reveal the nature of the underlying interaction. 
The dimension-6 operators have an additional dependence on the virtual photon's momenta in the Lagrangian, as shown in equations \ref{eq:lagrangianMDM}--\ref{eq:lagrangianCR}, and thus will display a strong difference in the kinematic distribution in the final state leptons, as is seen in the distribution of reconstructed photon momentum in the  bottom panel of Fig.~\ref{fig:sb}. Distributions for MDM and EDM models are very similar, as those from the CR and Anapole models; only CR and EDM are shown, for two choices of $m_T$.

\begin{figure}
    \centering
        \includegraphics[width=0.45\textwidth]{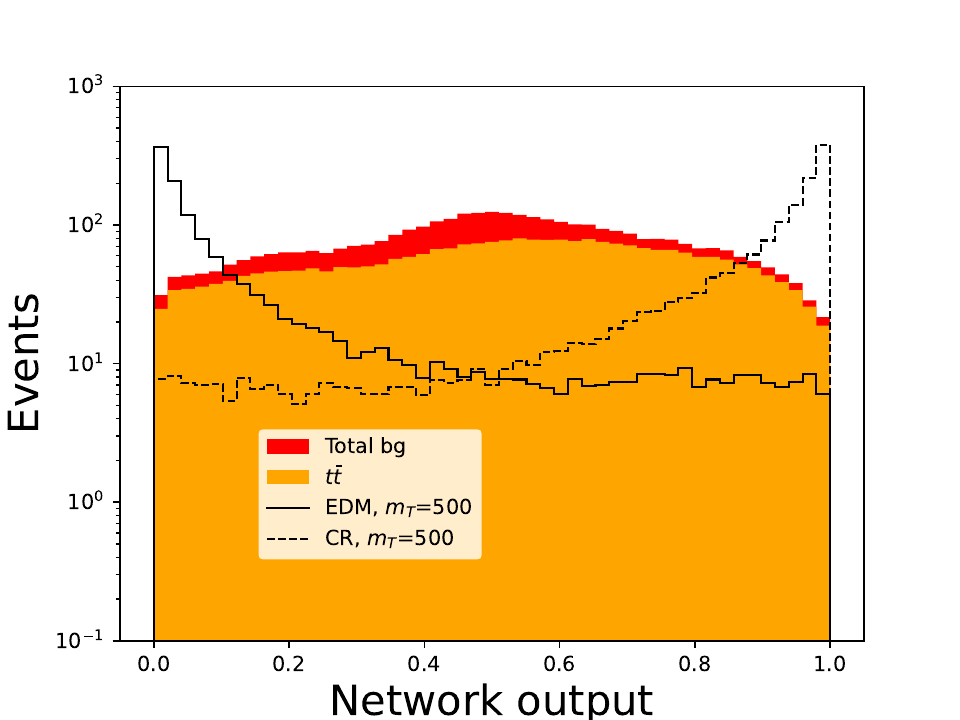}
        \includegraphics[width=0.45\textwidth]{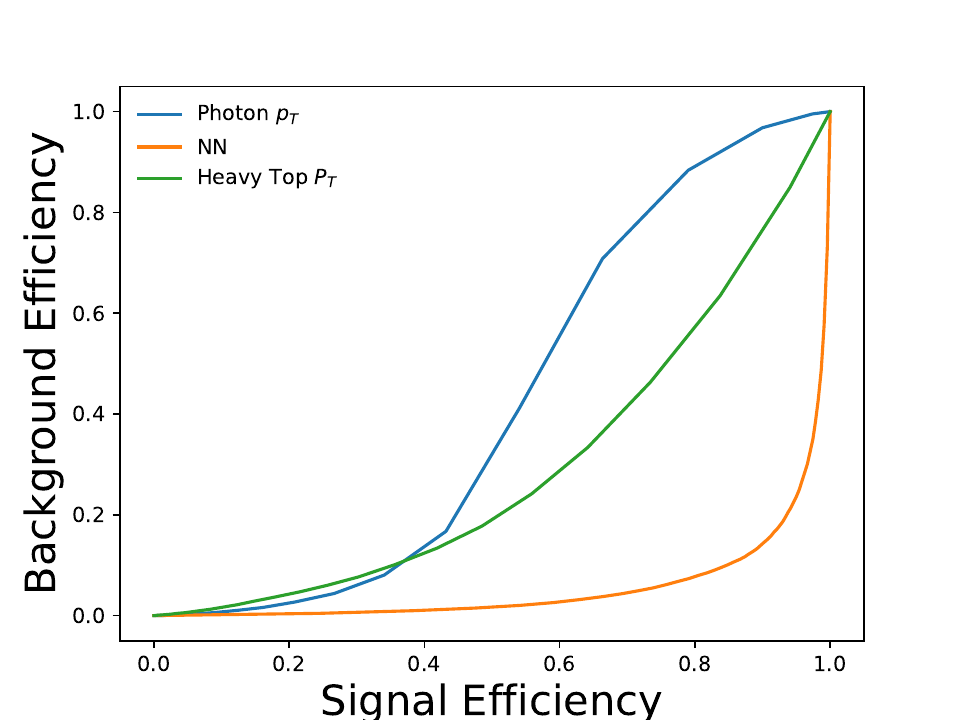}
                  \caption{Top: distribution of neural network output trained for the $m_T=500$ GeV scenario, shown for simulated signal and background samples normalized to an integrated luminosity of 300 fb$^{-1}$. The signal cross section is set to the expected limit at 95\% CL. Bottom: Signal and background efficiency for various thresholds on  photon transverse momentum, heavy quark transverse momentum or the neural network output, also for the $m_T=500$ GeV case.}
    \label{fig:nnperf}
\end{figure}

To capture the full discrimination information available, we train a neural network (NN) to distinguish between the heavy quark models. The network inputs are the four vectors of the reconstructed top quark and photon. Three hidden layers are used, consisting of 100 nodes, 80 nodes, and  20 nodes respectively, each with a ReLU activation. The final layer has a single sigmoid neuron.  In training, the dimension-5 models were given label 0 and the dimension-6 models label of 1. The loss function is binary cross entropy and mini-batch training with batch sizes of 10 was used,  along with the Adam optimizer. Each sample had approximately 100,000 events, and an independent network was trained for each of the candidate heavy quark masses considered.   The output of the network and its discrimination power is shown in Fig.~\ref{fig:nnperf}, showing significantly more discrimination power than the top or photon transverse momenta individually.

\begin{figure}
     \centering
     \scalebox{0.44}{\includegraphics{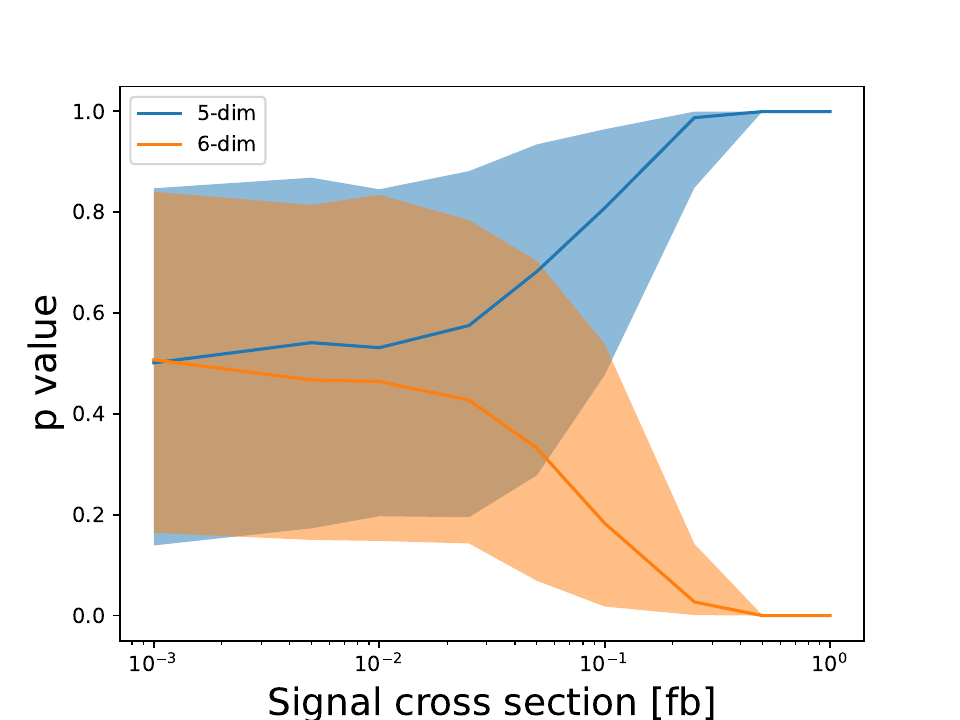}}
    \scalebox{0.44}{\includegraphics{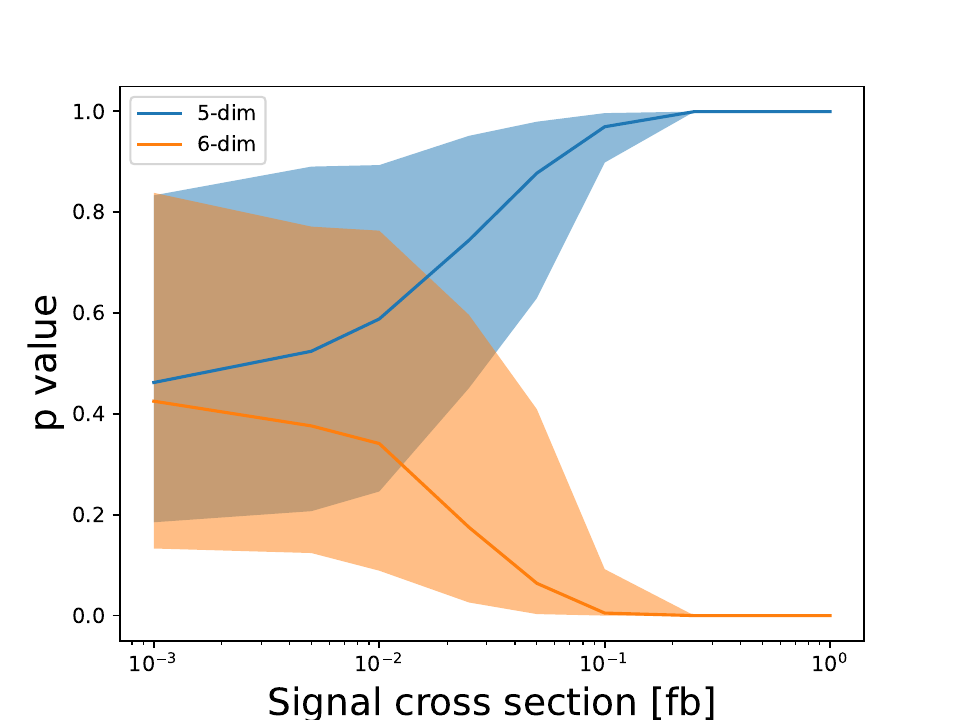}}
     \scalebox{0.44}{\includegraphics{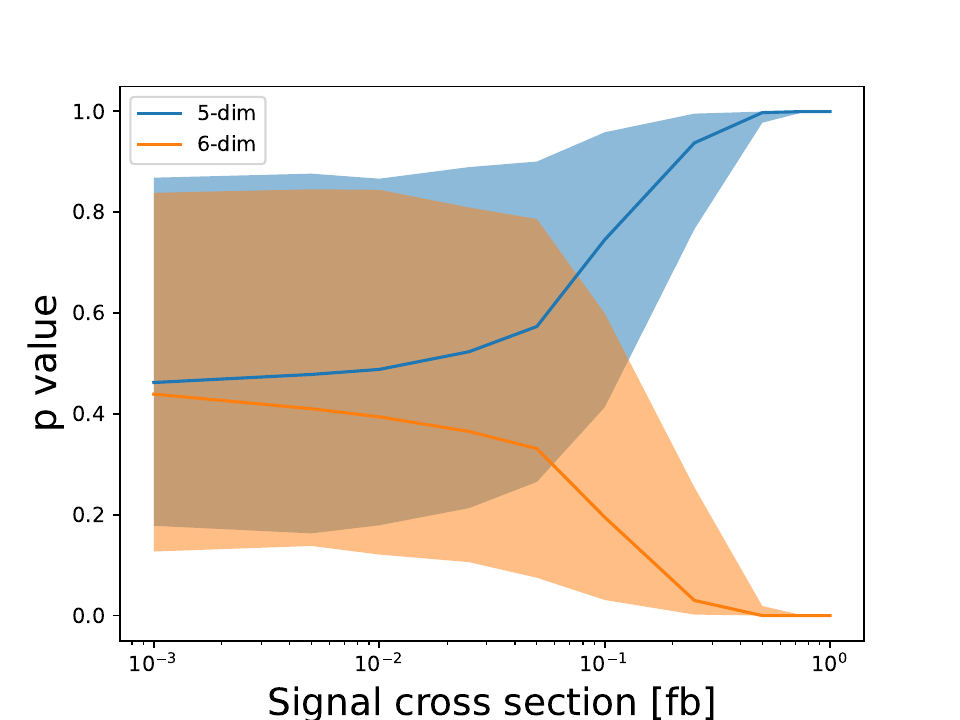}}
     \scalebox{0.44}{\includegraphics{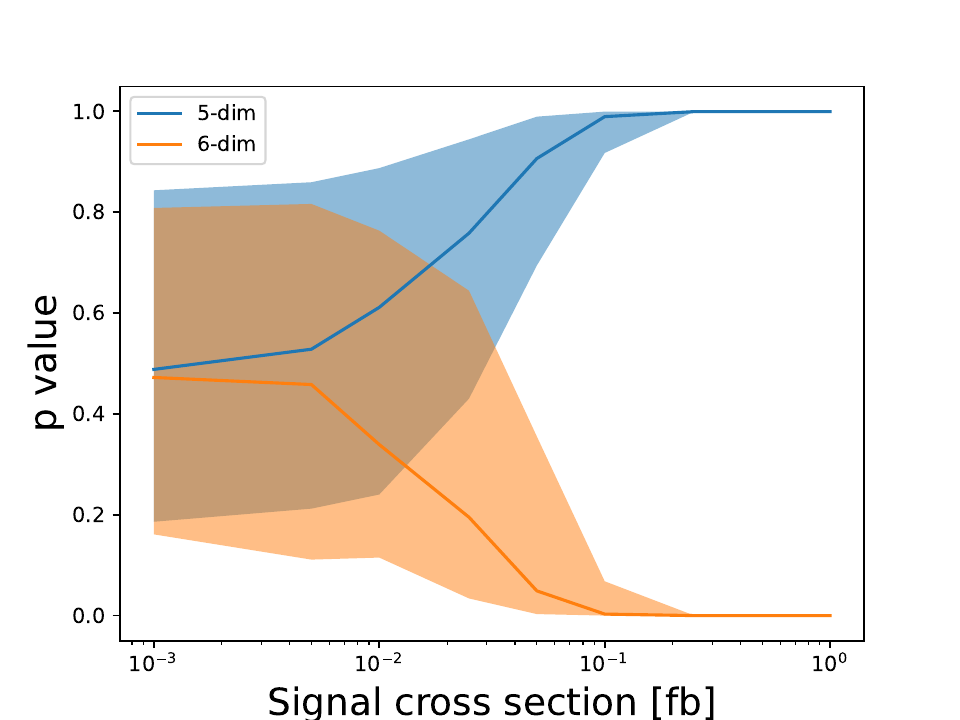}}
     \caption{ Expected $p$-values under a dimension-5 hypothesis of observing a NN output distribution expected from the dimension-6 hypothesis, and vice versa, as a function of the signal production cross section. Shown are the median case (solid line) and an envelope containing 95\% of hypothetical similar experiments for the       500, 1000, 1500 and 2000 GeV scenarios(top to bottom).}
     \label{fig:scans}
 \end{figure}

To assess the power of the dataset to distiguish between these models, we calculate the likelihood ratio (LR) between the dimension-5 and dimension-6 models using the binned NN output to estimate the densities. We calculate the $p$-value under the dimension-5 scenario of observing the mean LR expected in the dimension-6 scenario as a function of the signal cross section, and vice versa.  Results are shown in Fig.~\ref{fig:scans} and indicate that as the heavy quark mass increases, the signal cross section needed to obtain a $p$-value of less than 0.05 decreases, due to falling background rates. For $m_T \approx 1000$ GeV, sufficient statistics can be achieved for 300~fb$^{-1}$, for a cross section greater than about 0.1~fb. 

\section{Discussion}
\label{sec:disc}

While the various models produce the same list of final state objects and very similar heavy quark mass distributions, the full kinematics of the events allow us to disentangle the dimension-5 and dimension-6 hypotheses.   There are additional handles that might help, such as searching for on-shell production of photons, which are expected in dimension-5 models but not in dimension-6.   The on-shell photon in dimension-5 models also allow for a photon in the initial state, where $\gamma q\rightarrow T \rightarrow t\gamma$ gives a very similar final state and could contribute to the production rate, though the photon parton distribution function (PDF) is more uncertain than quark PDFs.

Other experiments and final states can also play an important role. In the event of non-zero mixing between $T$ and $u_R$, electroweak precision observables can be used to constrain mixing parameters \cite{Chen_2014,Chen_2017,Dawson_2012,Choudhury:2001hs}. Furthermore,  flavor changing neutral current processes have been used to constrain vector-like fermions \cite{Ishiwata_2015,Aaboud_2018,DELAGUILA1985243,Kang_2019,Balaji:2021lpr}. In addition to direct searches at the LHC, vector-like quarks in the ${\cal O}(100)$ GeV mass range can be searched for at the Tevatron~\cite{Okada:2012gy}.

The studies above demonstrate the possibility of distinguishing between dimension-5 and dimension-6 electromagnetic form factor models, but do not explore whether data could be used to prefer one of the models of equal dimensionality. The kinematics of the production and decay appear to be indistinguishable upon inspection of the distributions and attempts to train a machine learning model.

While our methods do not distinguish between operators of the same dimension, it may be possible to differentiate these models by leveraging observables related to angular correlations of the final state particles, which are sensitive to the CP structure of the interaction~\cite{Goncalves:2018agy,Goncalves:2021dcu}. Theoretically, distinguishing between parity-conserving and parity-violating operators of the same dimension could be achieved using a polarized beam, an option not available at the LHC, but potentially feasible at a future electron~\cite{Blondel:2019jmp} or muon collider~\cite{Norum:1996mi}. While these models do predict differences in the overall production rate at the LHC, the unknown coupling strength means that an observed rate could be compatible with either scenario. 

\section{Conclusions}
\label{sec:conc}

We present an example of a set of theoretical models which have different underlying mechanisms but would produce similar excesses in reconstructed invariant mass peaks. The theoretical distinctions lead to differences in other kinematic information in the event, which can be exploited in this case to disentangle two categories of  models, those with dimension-5 and dimension-6 operators.  We train a neural network to capture and summarize these differences. In a $pp$ collision dataset at $\sqrt{s}=13$ TeV with 300 fb$^{-1}$ and $m_T \approx 1000$ GeV, signal cross sections of approximately 0.1 fb are required to provide sufficient statistics to distinguish between the various hypotheses at 95\% CL.

\section{Acknowledgments} DW were funded by the DOE Office of Science. The work of M.Fieg was supported by NSF Grant PHY-2210283 and was also supported by NSF Graduate Research Fellowship Award No. DGE-1839285.  The authors thank Tim Tait, Avik Roy, Jeong Han Kim and KC Kong for useful comments.

\clearpage

\bibliography{stops}

\end{document}